\documentclass[aip,pof,reprint,groupedaddress]{revtex4-1}

\usepackage{graphicx}
\usepackage{float}
\usepackage{bm}
\usepackage{amsmath}
\usepackage{amssymb}
\usepackage{amsfonts}
\usepackage{amsbsy}
\usepackage{color}
\usepackage{txfonts}
\usepackage{hyperref}

\newcommand{\St}{\mbox{\textit{St}}}
\newcommand{\Rep}{Re_\mathrm{s}}

\begin{document}

\title{Concentrations of inertial particles in the turbulent wake of
  an immobile sphere}
\newcommand{\oca}{Laboratoire J.-L. Lagrange UMR\,7293, Universit\'e de
  Nice-Sophia Antipolis, CNRS, Observatoire de la C\^ote d'Azur, Bd.\
  de l'Observatoire, 06300 Nice, France}

\author{Holger \surname{Homann}}
\affiliation{\oca}
\author{J\'er\'emie \surname{Bec}}
\affiliation{\oca}

\date{\today}

\begin{abstract}
  Direct numerical simulations are used to study the interaction of a
  stream of small heavy inertial particles with the laminar and
  turbulent wakes of an immobile sphere facing an incompressible
  uniform inflow. Particles that do not collide with the obstacle but
  move past it, are found to form preferential concentrations both in
  the sphere boundary layer and in its wake. In the laminar case, the
  upstream diverging flow pattern is responsible for particle
  clustering on a cylinder that extends far downstream the sphere. The
  interior of this surface contains no particles and can be seen as a
  shadow of the large obstacle.  Such concentration profiles are also
  present in the case of turbulent wakes but show a finite
  extension. The sphere shadow is followed by a region around the axis
  of symmetry where the concentration is higher than the average. It
  originates from a resonant centrifugal expulsion of particles from
  shed vortices.  The consequence of this concentration mechanism on
  monodisperse inter-particle collisions is also briefly
  discussed. They are enhanced by both the increased concentration and
  the presence of large velocity differences between particles in the
  wake.
\end{abstract}

\maketitle

\section{Introduction}
Industrial and environmental problems often require modeling the
hydrodynamic interactions between particles suspended in a turbulent
flow. Sometimes it is necessary to accurately model the collision
efficiencies between droplets. Instances are rain triggering by
droplet coalescences in warm clouds\cite{pruppacher-klett:1997} or
planet formation by dust accretion in circumstellar
disks.\cite{lissauer:1993} In such problems one encounters different
regimes: For very tiny particles associated to small values of the
Reynolds number, the flow surrounding the particles is purely viscous
and evolves according to the Stokes equation. When dealing with a
large number of particles, this approach leads to model the collective
hydrodynamic interactions in terms of \emph{Stokesian
  dynamics}.\cite{brady-bossis-1988} In turbulent carrier flows, one
then generally approximates the flow as the superposition of the
small-scale effects of particles and the inertial-range Navier-Stokes
dynamics.\cite{wang2009turbulent} However, for larger particles with
non-negligible Reynolds numbers, the boundary layer and wake can
become rather complicated, so that for example even the drag force
experienced by such particles is up to now only heuristically modeled.
Therefore very little is known on the challenging problem of modeling
the hydrodynamic interactions between two particles with Reynolds
numbers implying that at least one particle has a turbulent wake.

\begin{figure}[t]
  \begin{center}
    \includegraphics[width=1\columnwidth,height=1.5cm]{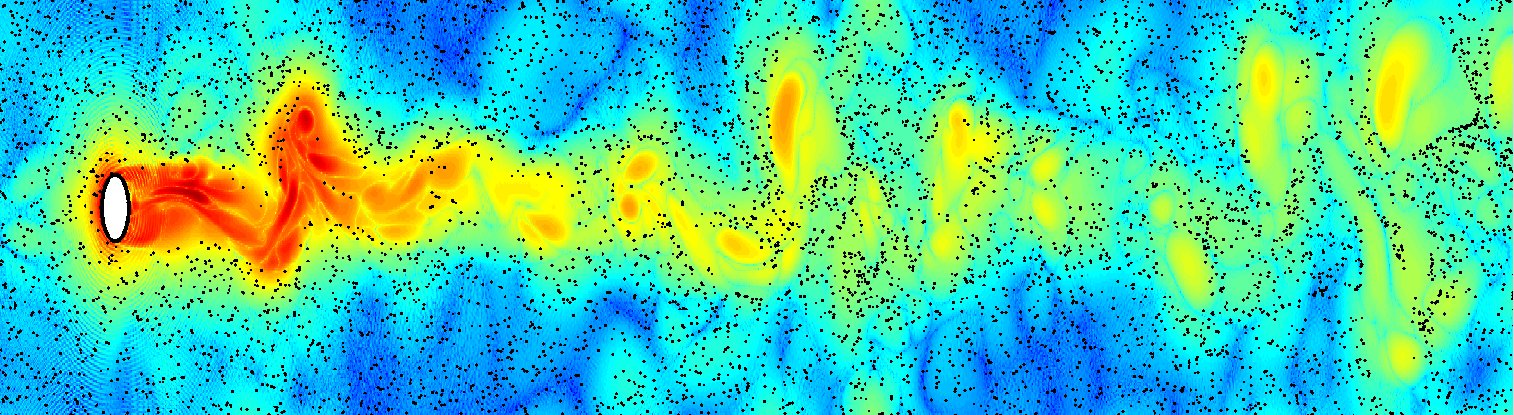}
  \end{center}
  \vspace{-27pt} $\ \xrightarrow{\displaystyle\quad U_C\quad }$\hfill\strut
  \caption{\label{fig:okuboWeiss_scatter} Snapshot of the position of
    particles (with $\St = \tau U_C/d = 3.2$, where $U_C$ is the
    homogeneous inflow velocity and $d$ is the diameter of the
    spherical obstacle) in a small slice passing through the axis of
    symmetry. The sphere Reynolds number is here
    $\Rep= U_C\,d/\nu = 400$. The background shows the values of the
    discriminant
    $\Delta = (\det[\mathbb{\sigma}]/{2})^2 -
    (\text{tr}[\mathbb{\sigma}^2]/{6})^3$
    which is positive in rotating regions and negative in stretching
    regions.}
\end{figure}

In this work we focus on the asymptotic case of particles with very
different sizes and investigate how a fixed sphere developing either a
laminar or turbulent wake interacts with a stream of very small heavy
particles. Inertia makes the trajectories of the small heavy particles
deviate from fluid element trajectories and this possibly leads to
particle-sphere collisions. One question is then to understand the
rate at which such collisions occur and how it is accelerated or
depleted as a function of the obstacle Reynolds number. This problem
has important applications in the determination of dust accretion by
planetesimals in early stellar systems or in the wet deposition of
tiny aerosols by raindrops and is studied elsewhere.  In the present
work, we are concerned with the dynamics of those particles which pass
the obstacle and interact with its wake. Using direct numerical
simulations, we study their spatial distribution and dynamics
downstream the spherical obstacle .

It is well known, that inertial particles do not exactly follow the
trajectories of fluid elements and distribute with an inhomogeneous
density.\cite{squires-eaton:1991,wood-hwang-etal:2005,bec-biferale-etal:2007}
Heavy particles are ejected from coherent rotating structures of the flow
by centrifugal forces and concentrate in stretching regions. The
creation of concentrations and voids is correlated to the local
structure of the fluid velocity gradient tensor; particles concentrate
in regions where it has real eigenvalues and voids appear in places
where two eigenvalues are complex.\cite{bec2005multifractal}
Figure~\ref{fig:okuboWeiss_scatter} shows a snapshot from one of our
simulations displaying small inertial particles in the turbulent wake
of a sphere. Detached coherent vortices that are advected downstream
by the mean flow eject particles and create voids. At the border of
these voids, clumps of particles can be observed. These clusters
trigger collisions that might be important in applications. Cloud
water droplets can be this way concentrated by a falling raindrop,
eventually coalesce, grow in size and influence the rain formation
process.

Tang \textit{et al.}\cite{tang-wen-etal:1992} and Yang \textit{et
  al.}\cite{Yang-crowe-etal:2000} studied numerically and
experimentally the influence of advected large scale structures on
particle dispersion in plane mixing layers and wakes. They show that
the development of highly organized patterns of inertial particles is
determined by the evolution and interaction of these structures. Via
the measurement of the fractal correlation dimension they found that
particles with a certain mass are the most influenced. Our work is
also related to an experimental study of Jacober and
Matteson.\cite{jacober-matteson:1990} In a wind tunnel they measured
average concentrations of inertial particles in the turbulent wake of
a sphere with Reynolds numbers ranging from 23000 to 110000. Despite
the fact that these Reynolds numbers are out of reach by direct
numerical simulations we find good qualitative agreement with their
data.

This article is organized as follows.
Section~\ref{sec:numericalMethod} gives a short description of the
numerical method applied. Section~\ref{sec:density} is dedicated to
the mean concentrations of small particles in the turbulent wake,
while Section~\ref{sec:twoParticle} presents an estimation of
inter-particle collisions. Section~\ref{sec:conclusion} encompasses
concluding remarks.

\section{The numerical method}
\label{sec:numericalMethod}

In order to conduct direct numerical simulations of a
three-dimensional hydrodynamic flow around a spherical object with
no-slip boundary conditions at its surface, we use a combination of a
standard pseudo-Fourier-spectral method with a penalty
method.\cite{homann-bec-etal:2013} More precisely, we solve the
incompressible Navier-Stokes equations
\begin{equation}
  \label{eq:navier-stokes}
  \partial_t {\bm u} + \bm u\cdot\nabla\bm u = -
  \frac{1}{\rho_\mathrm{f}}\nabla p +\nu \nabla^2 {\bm u} + {\bm f},
  \quad\nabla \cdot {\bm u} = 0,
\end{equation}
for the fluid velocity ${\bm u}$, where $\rho_\mathrm{f}$ is the fluid
density, $\nu$ its kinematic viscosity and ${\bm f}$ a force
maintaining a uniform and constant inflow. This mean flow is
imposed by keeping constant the zero mode of the stream-wise component
of the velocity; the mean velocity is denoted by $U_C$. Equation
(\ref{eq:navier-stokes}) is associated to the no-slip boundary
condition on the surface of the large particle at rest
\begin{equation}
  \label{eq:bc}
  {\bm u}(\bm x, t) = 0 \quad\mbox{for }\ |\bm x - \bm X_S| \leq {d}/2.
\end{equation}
Here $\bm X_S$ denotes the position of the particle center and $d$ its
diameter.  Homogeneous inflow conditions are achieved by using again
the penalty method to remove the velocity fluctuations originating
from the sphere wake at the exit of the computational domain, before
they are injected upstream by periodic boundary conditions. For the
time integration of (\ref{eq:navier-stokes}) we use a third order
Runge--Kutta scheme. The grid resolution is chosen to resolve all
small scales: those of the boundary layer of the sphere and all
turbulent scales in the wake. Details of the method and benchmarks can
be found in Homann \textit{et al.}\cite{homann-bec-etal:2013}

For the small inertial particles, we consider spherical particles with
a radius $a$ much smaller than both the viscous boundary layer and the
smallest active scale of the sphere turbulent wake (the local
Kolmogorov dissipative scale $\eta=(\nu^3/\epsilon)^{1/4}$, where
$\epsilon$ is the local energy dissipation rate). In this limit,
particles can be approximated by point particles. Further, we assume
that these particles move sufficiently slowly with respect to the
fluid and that their mass density $\rho_\mathrm{p}$ is much higher
than the fluid density $\rho_\mathrm{f}$. With these assumptions the
dominant hydrodynamic force exerted by the fluid is a Stokes viscous
drag, which is proportional to the velocity difference between the
particle and the fluid flow\cite{maxey-riley:1983,gatignol:1983}
\begin{equation}
  \ddot{\bm X} = \frac{1}{\tau}\! \left[\bm u(\bm X\!,t) \!-\!
  \dot{\bm X} \right], 
  \label{eq:particles}
\end{equation}
where the dots stand for time derivatives. The quantity $\tau =
2\rho_\mathrm{p} a^2/(9\rho_\mathrm{f} \nu)$ is called the particle
response time and is a measure of the particle inertia. It is the
typical relaxation time of the particle velocity to that of the
fluid. We also assume that the particles are sufficiently diluted to
neglect any interaction among them and any back-reaction on the
flow. Usually, particle inertia is measured in terms of the Stokes
number $\St = \tau\, / \, \tau_\mathrm{f}$ defined by
non-dimensionalizing their response time by a characteristic time
scale $\tau_\mathrm{f}$ of the carrier flow. The different time scales
involved here play a role depending on which aspect of the problem we
are interested in.  In most cases, we use $\tau_\mathrm{f} = d/U_C$,
which corresponds to expressing the particle response time in units of
the time needed to be swept over a distance $d$ by the large-scale
fluid velocity $U_C$. Particles with small $\St$ are closely
correlated to the flow and are swept around the obstacle. Large $\St$-
particles will preferentially collide with it. Only when concerned
with the particle dynamics in the turbulent wake we will use
$\St_\eta=\tau\, / \, \tau_\mathrm{f}$ with $\tau_\mathrm{f} =
\tau_\eta=(\nu/\epsilon)^{1/2}$ (the local Kolmogorov time scale).

We report results of three simulations corresponding to three
different values of the obstacle Reynolds number
$\Rep= U_C\,d/\nu = 100$, $400$, and $1000$. With this choice we cover
various wake types. For the smallest $\Rep$ the wake is steady and
laminar.  The intermediate value is slightly above the onset of
chaotic shedding of vortices and thus slightly turbulent. For
$\Rep=1000$ the wake is fully turbulent. In all cases, we consider
streams of heavy particles with response times $\tau$ between 0.04 and
40.96, corresponding to Stokes numbers in the range
$0.05\le\St\le63$. The main parameters of the simulations are
summarized in Tab.~\ref{table}.

\begingroup
\squeezetable
\begin{table*}[h]
  \centering
  \begin{ruledtabular}
  \begin{tabular}{cccccccccccrrc}
    $\Rep$ & $U_C$ & $d$ & $\nu $ & $\ell_x\times \ell_y\times
                                    \ell_z$  & $N_x\times N_y\times N_z$ & $N_p$ \\ \hline
    100  & 1  & 0.8  & $8\cdot 10^{-3}$ & $2\pi\times2\pi\times16\pi$
                                             & $256 \times 256 \times 2048$ &$\approx 10^6$
    \\
    400  & 1  & 0.8  & $2\cdot 10^{-3}$ & $2\pi\times2\pi\times16\pi$
                                             & $256 \times 256 \times 2048$ &$\approx 10^6$
    \\
    1000  & 1 & 0.65 & $6.5\cdot 10^{-4}$ &
                                            $2\pi\times2\pi\times16\pi$ & $512 \times 512
                                                                          \times 4096$ &$\approx 10^6$
  \end{tabular}
  \end{ruledtabular}
  \caption{\label{table}Parameters of the numerical simulations.
    $\Rep=U_C\,d/\nu$: Reynolds number of the obstacle, $U_C$: inflow
    velocity, $d$: diameter of the spherical obstacle, $\nu$: kinematic
    viscosity, $\ell_x$, $\ell_y$ and $\ell_z$: edge lengths of the computational
    domain ($z$ being in the direction of the mean flow), $N_x$,
    $N_y$, and $N_z$: number of collocation points in the respective
    directions, $N_p$: number of small particles.}
\label{table1}
\end{table*}
\endgroup
 
The simulations are set up in the following way. First, we integrate
the Navier-Stokes equations~(\ref{eq:navier-stokes}) with a sphere
located at $\bm X_S=(\pi,\pi,2\,\pi)$ while imposing a homogeneous
inflow with $U_c=1$ until a (statistically) stationary state is
reached. We then inject a stream of inertial particles by placing
randomly many particles per time step at the entrance of the
domain. They are initialized with the velocity $U_C=1 $ of the
fluid. During the simulation we remove all the particles that are
touching the sphere or reaching the end of the computational
domain. On average the domain is filled with approximately one million
particles. The fluid and particle trajectories are then stored and
analyzed during four times the time needed for the fluid to travel
across the numerical domain.

\section{Density inhomogeneities in the wake}
\label{sec:density}
\subsection{Average concentrations and voids}

\begin{figure}[h]
  \begin{center}
    \includegraphics[width=0.99\columnwidth]{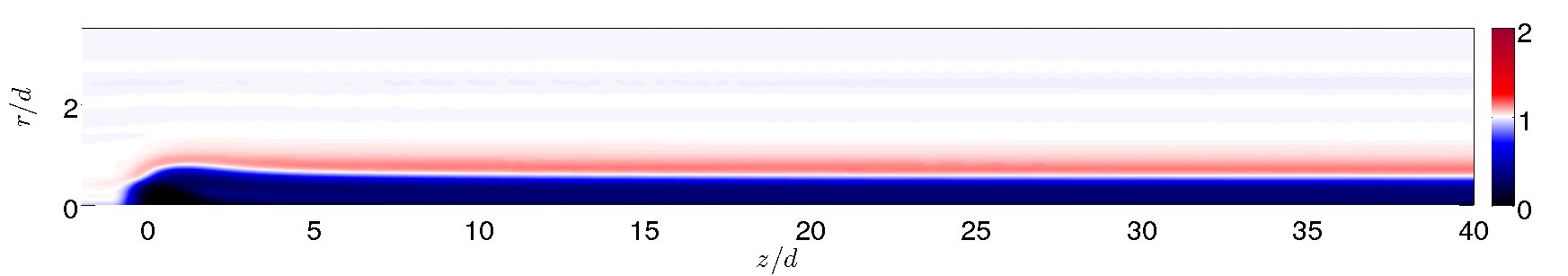} 
    \includegraphics[width=0.99\columnwidth]{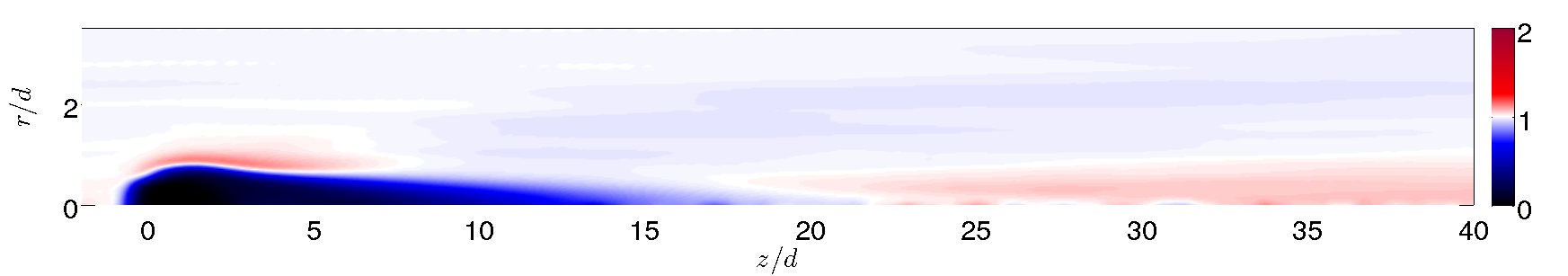} 
    \includegraphics[width=0.99\columnwidth]{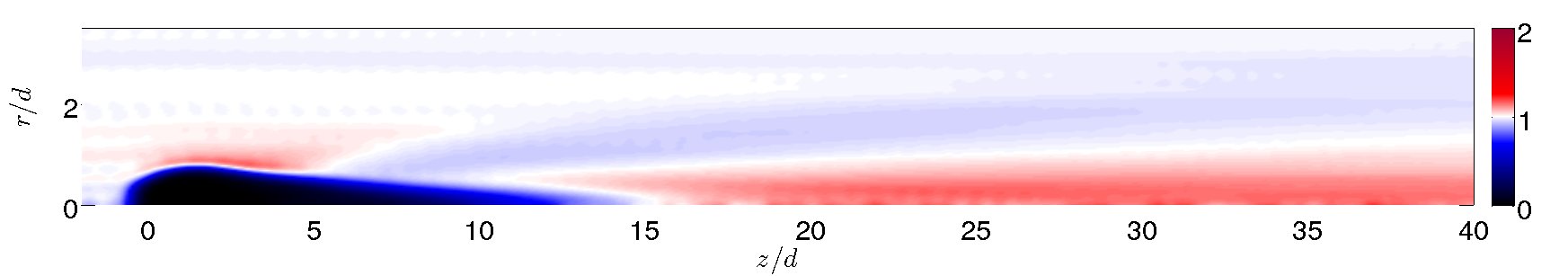} 
   \end{center}
   \caption{\label{fig:meanConcentration} Mean particle concentration
     for $\St=0.4$ and $\Rep=100$ (top), $\St=0.4$ and $\Rep=400$
     (middle) and $\St=0.49$ and $\Rep=1000$ (bottom).}
\end{figure}

Small inertial particles that pass over the spherical obstacle are
influenced by both its boundary layer and its wake. The rotating
structures in the wake eject particles and create voids and
concentrations (see again Fig.~\ref{fig:okuboWeiss_scatter}). The mean
(temporal average) particle number density is shown in
Fig.~\ref{fig:meanConcentration} for the three different Reynolds
numbers we have considered. Due to the axial symmetry of the problem,
we adopt a cylindrical coordinate system $(r,z,\phi)$ (where $z$ is in
the direction of the inflow $U_C$) and average over the angle
$\phi$. One observes regions where the particle concentration is
either significantly over or under its value in the inflow stream.

Let us outline the main findings: A cylindrical region where small
particles concentrate is emerging at the surface of the obstacle while
a shadow with a very low concentration extends downstream. In the case
of the two highest Reynolds numbers (turbulent wakes), this shadow
region is drastically reduced and, moreover, another
over-concentration is created further downstream.  This second
high-concentration region around the axis of symmetry is accompanied
by a conic region of reduced concentration spreading outwards. Both
the high as well as the low concentrations are more pronounced for the
higher Reynolds number. For $\St=3.9$ and $\Rep=1000$, we observe for
example an over-concentration in the wake that is approximately 1.5
times larger than the mean concentration.

\subsection{Stream-wise and radial profiles}

\begin{figure}[t]
  \begin{center}
    \includegraphics[width=0.49\columnwidth]{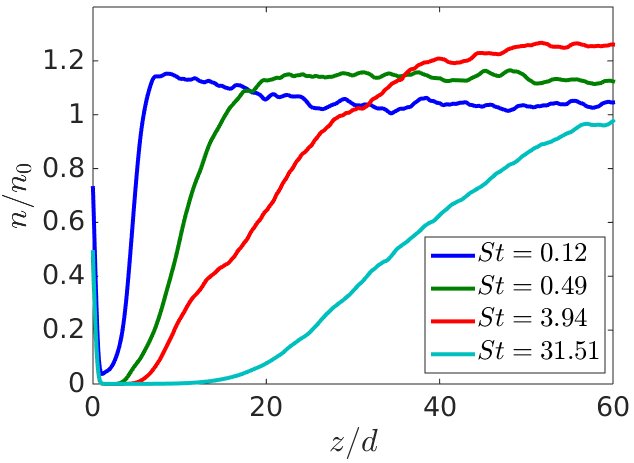}
    \includegraphics[width=0.49\columnwidth]{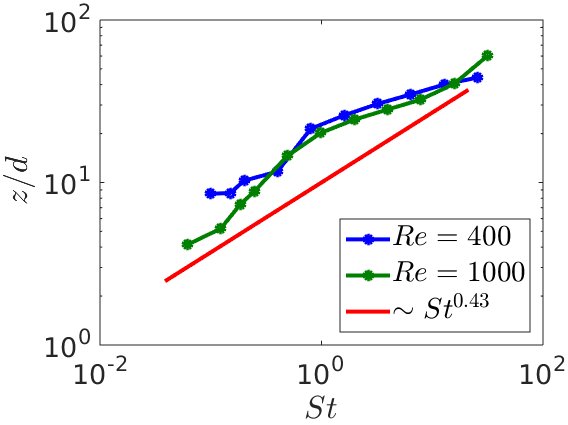}
   \end{center}
   \caption{\label{fig:meanConcentration_axis} Left: Mean particle
     concentration on the axis of symmetry as a function of the
     distance from the spherical obstacle for several $\St$ and
     $\Rep=1000$. Data has been smoothed by a sliding window average
     with a window size approximately equal to $d$. Right: Position in
     the wake where the small particle density recovers the inflow
     density, corresponding to the transition between the shadowed
     region and the over-concentration.}
\end{figure}

The spatial dependence of these concentrations can be seen with more
details in Fig.~\ref{fig:meanConcentration_axis} (left). Particles
with a small Stokes number $\St$ only create a short shadow (a few
times the obstacle diameter $d$ for $\St=0.12$) where the
concentration is significantly depleted. Rapidly, at only several
diameters further downstream, the density exceeds the inflow density
and reaches a maximum. Beyond this point, the over-concentration
slowly reduces. This is because $\tau_\eta$ increases (and therefore
$\St_\eta=\tau/\tau_\eta$ reduces) as a function of the distance from
the sphere so that small-$\St$ particles behave more and more like
tracers and stick to fluid elements. They are transported and mixed by
the wake turbulent fluctuations and this evens out possible
concentrations.

The density profile for particles with a larger Stokes number $\St$ is
similar but shifted further downstream due to higher inertia. The
concentrations might persist in the far wake as turbulence decays
downstream so that heavy particles do not diffuse anymore and their
density might just be advected by a quasi-uniform flow. Also, it seems
that the maximum of concentration increases with $\St$. However, these
two observations cannot be unquestionably confirmed from our
simulations, given the limited span in the streamwise direction. A
more precise investigation of these aspects would require a much
longer computational domain and is kept for a future study.

\begin{figure}[h]
  \begin{center}
    \includegraphics[width=0.55\columnwidth]{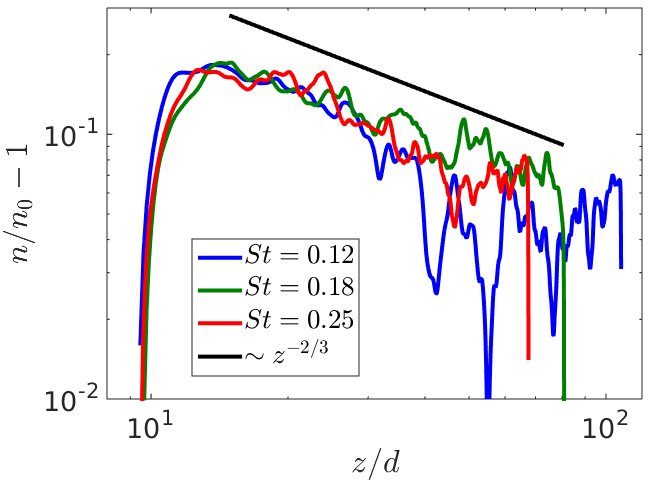}
  \end{center}
  \caption{\label{fig:meanConcentration_stream} Left: Mean particle
    concentration deficit on the axis of symmetry as a function of the
    distance from the obstacle center for several $\St$. Data has
    been smoothed by a sliding window average with a window size of
    approximately $d$. Data is shown for $\Rep=1000$. The black line
    corresponds to a decay $\propto z^{-2/3}$.}
\end{figure}

The starting point of the over-concentration region (where the
particle density exceed that of the inflow) is shown in
Fig.~\ref{fig:meanConcentration_axis} (right) as a function of
$\St$. It follows a power-law with an exponent which is close to that
measured by Jacober and Matteson.\cite{jacober-matteson:1990} The
decrease of the over-concentration as a function of the distance from
the obstacle is reported in
Fig.~\ref{fig:meanConcentration_stream}. Here, only small-$\St$
particle are considered as the position of the maximal density lies
outside the domain for large $\St$. The decay follows a power-law
with an exponent close to $-2/3$.

\begin{figure}[h]
  \begin{center}
    \includegraphics[width=0.48\columnwidth]{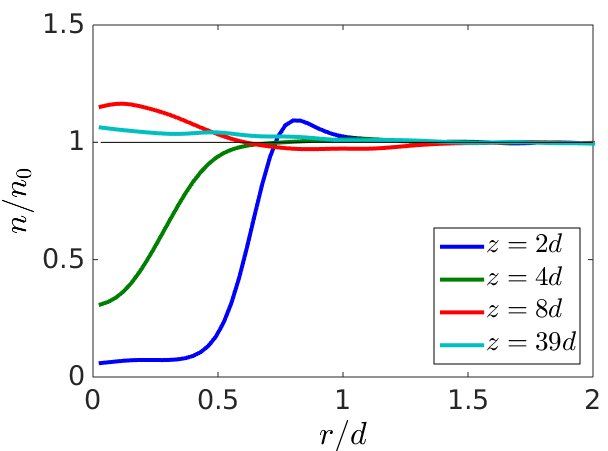} \hfill
    \includegraphics[width=0.48\columnwidth]{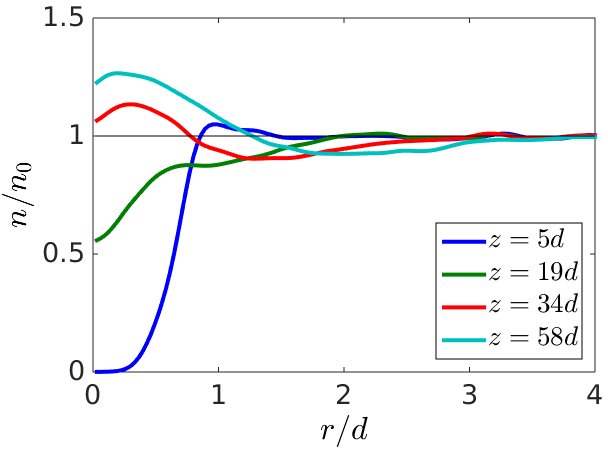}
  \end{center}
  \caption{\label{fig:trans} Radial density profiles for various
    distances downstream the spherical obstacle ($\Rep=1000$) and for
    $\St=0.12$ (left) and $\St=3.94$ (right)}
\end{figure}

The radial profiles of the particle density are shown in
Fig.~\ref{fig:trans} for two different Stokes numbers. At small
distances $z$ from the obstacle, the high-concentration jet emerges at
a radial distance $r\approx d$ from the axis of symmetry. Further
downstream, the transition from regions of reduced to increased
particle concentrations are also situated at $r\approx d$.

\subsection{Concentration mechanisms}

\begin{figure}[h]
  \begin{center}
    \includegraphics[width=0.85\columnwidth]{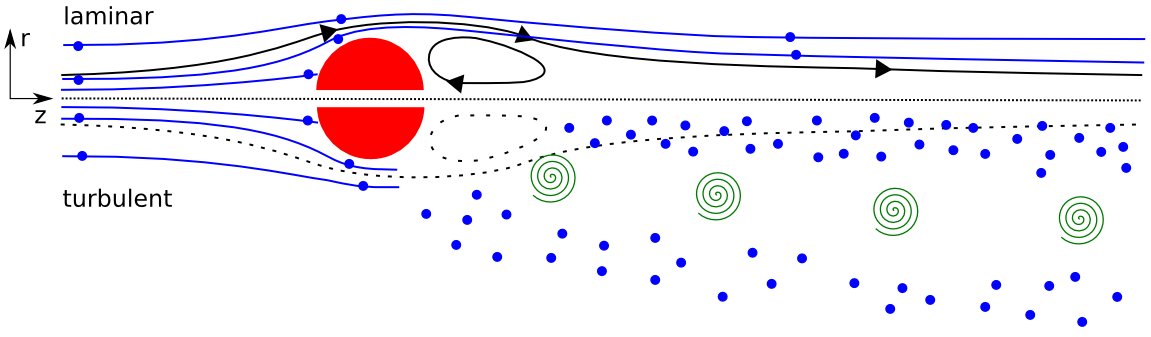}
  \end{center}
  \caption{\label{fig:concentrationSketch} Sketch showing the
    mechanisms leading to over- and under-concentrations of small
    particles in the wake of a spherical obstacle (red half-disk). The
    upper half shows the situation for laminar wakes while the lower
    part relates to turbulent wakes. Streamlines are shown in
    black. In the case of a turbulent wake, the stream lines of the
    mean flow are shown as dashed lines. Blue lines correspond to
    trajectories of heavy particles. In the turbulent wake particle
    positions are indicated by blue points. Green spirals sketch the
    position and rotation of detached vortices.}
\end{figure}

For all Reynolds numbers of the spherical obstacle, a
high-concentration cylindrical jet of particles emerges directly from
the surface of the sphere. The reason for this is the converging flow
in front of the obstacle (see the sketch in
Fig.~\ref{fig:concentrationSketch}). Particle trajectories that start
upstream at locations close to the axis of symmetry (small $r$'s) come
close to those starting with a larger $r$. This is due to reduced
accelerations of the inertial particles compared to fluid
elements. The observed number densities in the jet region decrease
with increasing Stokes number $\St$. This is because the collision
rate of small particles with the sphere increases with $\St$. The
particles which have collided are removed from the flow and are thus
missing downstream and in the jet region.

The low-concentration shadow behind the obstacle originates from the
fact that the dynamics of inertial particles lies behind that of the
fluid. Fluid streamlines converge downstream the recirculation region
but inertial particles overshoot. The resulting shadow persists far
away in the wake (in the laminar case), because of the turbulent decay
downstream and the associated decrease of the fluid radial
acceleration. Particles get less and less driven toward the center
axis of symmetry.

How can one understand the emergence of the over-concentration regions
in the far turbulent wakes? Jacober and
Matteson\cite{jacober-matteson:1990} suggested that they originate
from the jet region that is drawn around the sphere and collapses
behind it. However, if this was the case, a laminar wake would also
produce similar concentrations and we do not observe this effect. The
mean flow in the turbulent wake is not responsible for these
concentrations either. We indeed integrated trajectories of inertial
particles according to (\ref{eq:particles}) but transported by the
mean velocity field $\bar{\bm{u}}(\bm x)$, \textit{i.e.}\/ the
temporally averaged velocity field of the $\Rep=400$ simulation (not
shown). Up to several diameters downstream from the fixed sphere, the
concentration profile is very similar to the time dependent
simulations. Notably, the jet is reproduced. However, this approach
does not produce the high-concentration region in the far wake.

\begin{figure}[h]
  \begin{center}
      \includegraphics[width=0.99\columnwidth]{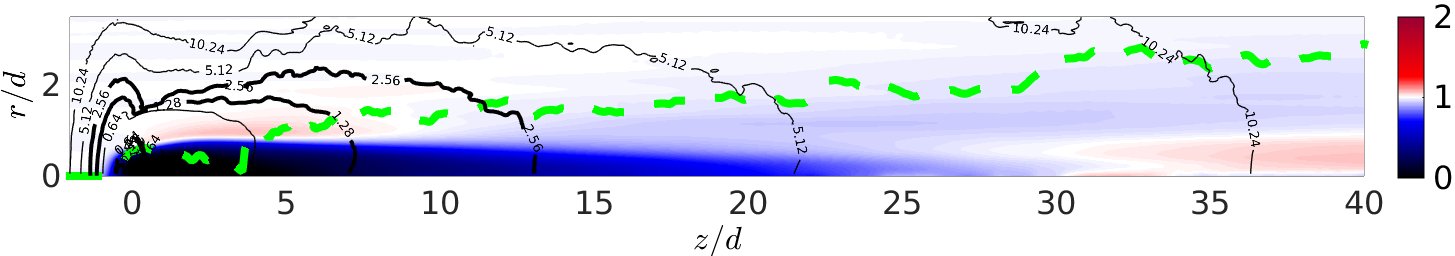}
  \end{center}
  \caption{\label{fig:re400_vr_maxOkubo} Mean particle concentration
    in pseudo colors for $\St=1.6$ ($\tau=1.28$) particles and
    $\Rep=400$. Position of the maximal $\Delta$ as a function of the
    downstream distance $z$ is shown as a dashed line. Solid black
    lines indicate contours of the local Kolmogorov time-scale
    $\tau_\eta$.}
\end{figure}

We suggest that the concentration mechanism is provided by the
rotating structures in the turbulent wake. Yao \textit{et
  al.}\cite{yao_zhao-etal:2012} already observed instantaneous
concentrations in a two dimensional flow past a disk and found that
particles concentrate between periodically shaded vortices. Tang
\textit{et al.}\cite{tang-wen-etal:1992} and Yang \textit{et
  al.}\cite{Yang-crowe-etal:2000} studied numerically and
experimentally the influence of advected large scale structure on
particle dispersion in plane mixing layers and wakes. They found that
the development of highly organized patterns of inertial particles is
determined by the evolution and interaction of these structures. Via
fractal correlation dimension measurements they showed that particles
with $\St=1$ are the most influenced.

In order to study with more details the effects of the wake rotating
structures on particle concentrations, we computed the temporally
averaged Okubo--Weiss parameter $\Delta$ (defined in the caption of
Fig.~\ref{fig:okuboWeiss_scatter}). It is a measure of the strength of
rotation \textit{vs.} stretching in the fluid velocity field. $\Delta$
is positive in regions where the velocity gradients are dominated by
vorticity and negative in those where strain is dominant. In a
turbulent flow, heavy inertial particles tend to move from regions
where $\Delta$ is positive to those where it is
negative.\cite{bec-biferale-etal:2007} The maximum of $\Delta$ as a
function of the streamwise position is shown as the dashed line in
Fig.~\ref{fig:re400_vr_maxOkubo}. One clearly observes that it
correlates well with the region of under-concentration.

Another observation is that, when the particle inertia is increased,
the starting point of the concentration region is shifted downstream
(compare Fig.~\ref{fig:meanConcentration_axis}).  This change can be
explained by the fact that particle ejection from rotating regions is
the most effective for particles with a Stokes number $\St_\eta =
\tau/\tau_\eta$ around unity.  The local Kolmogorov time scale
$\tau_\eta$ increases downstream (see solid black contour lines in
Fig.~\ref{fig:re400_vr_maxOkubo}). Hence, the position where
$\St\approx 1$ shifts downstream when particle inertia increases.

\section{Inter-particle collisions}
\label{sec:twoParticle}
The goal of this section is to qualitatively estimate the occurrence
and position of possible collisions between particles with equal
inertia. We will limit the discussion to the flow $\Rep=400$ and we
focus on mean quantities, leaving the influence of fluctuations for a
future work. Based on the mean pair density and on the mean
approaching velocity between particles with the same Stokes number,
one can estimate the collision rate is proportional to the product
\begin{equation}
p_c(z,r) = -\langle n_-(z,r)^2 \rangle \, \langle w_-(z,r) \rangle,
\label{eq:def_pc}
\end{equation}
where $n_-^2$ is the coarse-grained pair density of approaching
particles and $w_-$ the associated mean longitudinal velocity
difference. For both $n_-^2$ and $w_-$, only particles with negative
longitudinal velocity differences are taken into account.  Such an
estimate for collision rates relies on the ghost-particle
approximation\cite{sundaram-collins:1997} where possible particle
encounters are counted but not performed. It relies on providing an
arbitrary radius to the particles that we have here chosen to be
$0.2\,d$. All the coarse-grained quantities used in (\ref{eq:def_pc})
are thus defined for that scale. We observe from our simulations that
the pair density $\langle n_-^2 \rangle$ has the same structure as the
mean single particle concentrations (not shown): A high concentration
jet region in the vicinity of the large sphere for particles with a
small inertia and a high concentration region downstream, more
pronounced for particles with a large inertia.

\begin{figure}[h]
  \begin{center}
    \includegraphics[width=0.99\columnwidth]{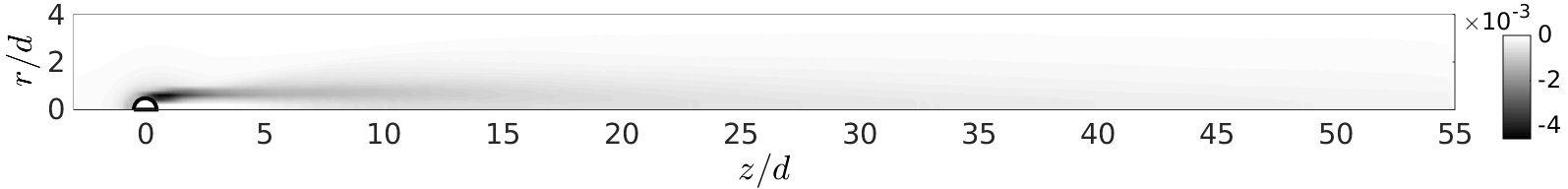} \\
    \includegraphics[width=0.99\columnwidth]{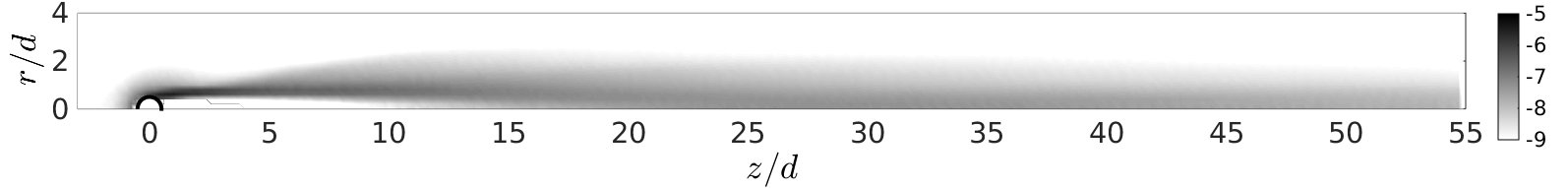} 
  \end{center}
  \caption{\label{fig:w2} Top: Average velocity difference $\langle
    w_-(z,r) \rangle$ of approaching particles with $\St=3.2$ ,
    coarse-grained at a scale $0.2d$. Bottom: Estimate $p_c$, defined in
    (\ref{eq:def_pc}), for binary collisions of particles with the
    same $\St=3.2$. It is represented in logarithmic units in order
    to emphasize its variation rather than only its maximum in the
    jet region.}
\end{figure}

The average longitudinal velocity differences are shown in
Fig.~\ref{fig:w2} (top). The highest approaching velocities are
located in the jet region, where particle trajectories are
compressed. In the wake region where vortices detach, one still
observes some non-vanishing values of the approaching velocities,
which are much less strong than in the jet region. Further in the
wake, the velocity differences become weak due to the decay of
turbulent fluctuations.

\begin{figure}[h]
  \begin{center}
    \includegraphics[width=0.99\columnwidth]{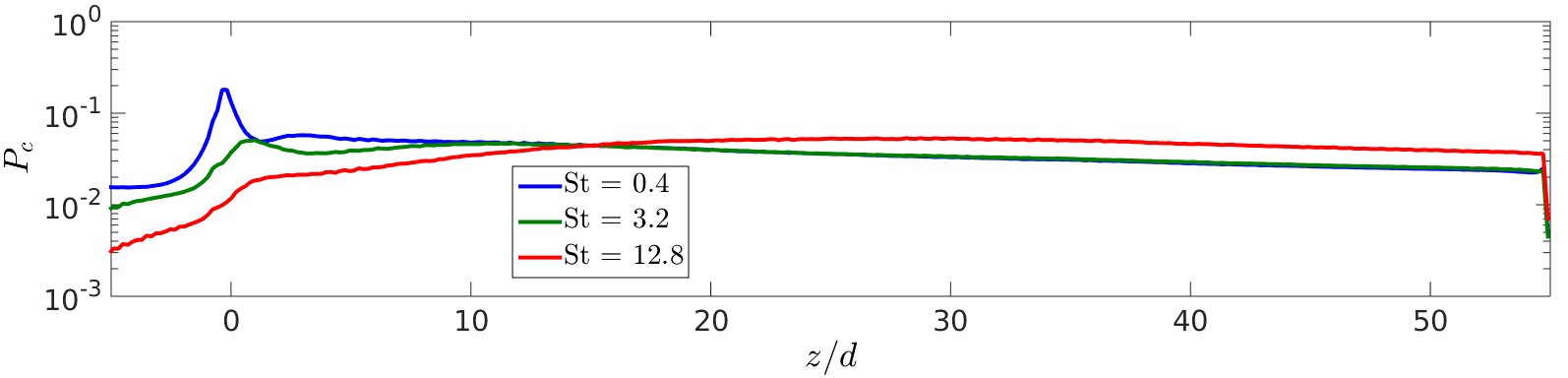} 
  \end{center}
  \caption{\label{fig:n2w2sum} Sum of $p_c$ over planes ($r,\theta$)
    in directions perpendicular to the stream, as a function of the
    downstream distance to the obstacle. Three different strengths of
    particle inertia are shown, as labeled. }
\end{figure}

The estimate $p_c$ for collision rates is shown in
Fig.~\ref{fig:w2} (bottom). The jet region provides the highest probability
for particle encounters. There, both the particle density as well as
the particle approaching speed are enhanced. With increasing distance
from the obstacle the particle collision region moves towards the axis
of symmetry and the probability of collisions decreases. The
dependence of the collision probability as a function of the distance
from the obstacle can be seen more clearly in Fig.~\ref{fig:n2w2sum},
where we show the sum of $p_c$ over planes ($r,\theta$) as a function
of the streamwise direction $z$. One clearly observes that the
particles with a small inertia collide preferentially in the jet
region, in the vicinity of the sphere. The collisions between
particles with a larger inertia are rather enhanced in the wake. We
indeed see in Fig.~\ref{fig:n2w2sum} that non-vanishing values of
$p_c$ persist at downstream distances as far as tens of the obstacle
diameter.

\section{Summary and outlook}
\label{sec:conclusion}
In this work we have analyzed how a stream of inertial particles is
influenced by the boundary layer and the wake of a spherical
obstacle. We make use of the results of three direct numerical
simulations of the incompressible Navier--Stokes equations using a
combination of a Fourier-spectral method and a penalty technique.  We
considered three different Reynolds numbers for the sphere,
$\Rep=100$, $\Rep=400$, and $\Rep=1000$. The average concentration of
small inertial particles shows different characteristics in the wake
of the spherical obstacle, depending on the particle Stokes number
$\St$ and on the sphere Reynolds number $\Rep$: For all $\Rep$ we find
that particles with weak or moderate inertia cluster at the outer edge
of the obstacle creating a narrow cylindrical jet of particles. Behind
the obstacle, inertia is responsible for the formation of a shadow
where particle concentration is strongly reduced.  When the wake of
the obstacle is laminar, these over- and under-concentrations persist
far downstream while they are rapidly dissipated by fluctuations for a
turbulent wake. Moreover, we find that in the turbulent case, there is
a region appearing a couple of diameters downstream the sphere where
the particle concentration exceeds that in the inflow. This can be
explained by resonant ejections of particles from detached and
advected vortices with a local turn-over time $\tau_\eta$ of the order
of the particle response time $\tau$.

Estimates of the probability of collisions between identical particles
show that particles with a weak inertia tend to collide in the jet
region near the sphere while those with a strong inertia collide
preferentially further downstream. In a future work it will be
interesting to examine such collision statistics in the polydisperse
case, \textit{i.e.}\/ between particles with different sizes and
response times. We expect that in the turbulent wake the relative
velocity of particles with different inertia will be larger than for
particles of the same type. We also plan to take coagulation or
coalescences into account and analyze how the particle size
distribution is affected by the wake of the spherical obstacle and how
it depends on the Reynolds number of the latter and on the downstream
distance. This can have important applications, for instance in planet
formation. It seems well founded to assume that large bodies, such as
planetesimals or planetary embryos, will concentrate dust in their
wake and will thus foster accretion and the formation of other
objects.
\subsection*{Acknowledgments}
We thank P.~Tanga and T.~Guillot for useful discussions. Most of the
simulations were done using HPC resources from GENCI-IDRIS (Grant
i2011026174). Part of them were performed on the âm\'{e}socentre de
calcul SIGAMMâ. The research leading to these results has received
funding from the Agence Nationale de la Recherche (Programme Blanc
ANR-12-BS09-011-04).

\bibliographystyle{ieeetr}

\end{document}